# MRANet: A Modified Residual Attention Networks for Lung and Colon Cancer Classification

Diponkor Bala*, S M Rakib Ul Karim†, Rownak Ara Rasul†
*Department of Computer Science and Engineering, City University, Savar, Bangladesh
†Department of Electrical Engineering and Computer Science, University of Missouri, Columbia, MO, USA
Email: diponkor@cityuniversity.ac.bd, skarim@missouri.edu, rrasul@missouri.edu

*Abstract*—Lung and colon cancers are predominant contributors to cancer mortality. Early and accurate diagnosis is crucial for effective treatment. By utilizing imaging technology in different image detection, learning models have shown promise in automating cancer classification from histopathological images. This includes the histopathological diagnosis, an important factor in cancer type identification. This research focuses on creating a high-efficiency deep-learning model for identifying lung and colon cancer from histopathological images. We proposed a novel approach based on a modified residual attention network architecture. The model was trained on a dataset of 25,000 high-resolution histopathological images across several classes. Our proposed model achieved an exceptional accuracy of 99.30%, 96.63%, and 97.56% for two, three, and five classes, respectively; those are outperforming other state-of-the-art architectures. This study presents a highly accurate deep learning model for lung and colon cancer classification. The superior performance of our proposed model addresses a critical need in medical AI applications.

## I. INTRODUCTION

Lung and colon cancers rank among the most common and lethal malignancies worldwide, with millions of new cases documented each year. Timely and precise diagnosis is crucial for enhanced therapy efficacy and better patient outcomes. There may be inconsistency in diagnosis since traditional methods of detecting malignant cancers, such as biopsy and histological examination, are laborious and mostly rely on the expertise of pathologists. Histopathological image-based identification of lung and colon cancer still faces significant challenges due to high variability in tissue morphology, staining, and imaging conditions, which may introduce large discrepancies across datasets [1].

In order to facilitate quicker and more reliable diagnosis, advances in artificial intelligence (AI), particularly in the area of deep learning, have demonstrated significant promise in automating the categorization of malignant tissues from medical imaging [2]. In addition, deep learning models are often not interpretable, which limits their clinical adoption. A very promising avenue for overcoming these challenges is the residual attention network model. This model has been developed on the basis of the residual connections and attention mechanisms that help it focus on critical areas of the complex histopathological images by capturing local and global features while preserving the flow of information, hence increasing its accuracy and interpretability and possibly allowing more robust and interpretable cancer detection on lung and colon tissue samples [3].

This work will introduce a new model based on the modified residual attention network that can be applied to the task of lung and colon cancer categorization. This work thus aims at building an efficient model with high accuracy, which can be accepted by medical professionals, through lightweight residual learning network designs combined with robust feature extraction and attention techniques. Building on recent advancements in medical imaging and artificial intelligence, this work aims to help ongoing efforts to improve cancer diagnosis and treatment through technology, ultimately leading to better patient outcomes [4]. The major contributions of this study are:

- The suggested modified residual attention network model, termed "MRANet", demonstrated superior performance compared to existing models in classifying lung and colon cancers, with high accuracies of 99.30%, 96.63%, and 97.56% for two, three, and five classes, respectively.
- Our approach is rigorously assessed on the employed benchmark histopathology dataset, exhibiting superior performance relative to current state-of-the-art techniques.
- Our work has the potential to assist pathologists in improving diagnostic accuracy and consistency, contributing to better clinical outcomes for lung and colon cancer patients.

The subsequent sections of the paper are structured as follows: Section II presents the pertinent literature review. Section III delineates the approach, encompassing dataset preparation, and the recommended model designs. Section IV will present and analyze our findings and debate the results obtained. Ultimately, Section V ends the work by encapsulating the primary contributions and proposing directions for future research.

## II. RELATED WORKS

Significant advancements have been achieved in identifying the types of lung and colon cancers through the use of histopathology images, driven by the implementation of deep learning, which has notably improved both efficiency and accuracy. For conducting this study, we have studied some related previous paperwork that is discussed below.

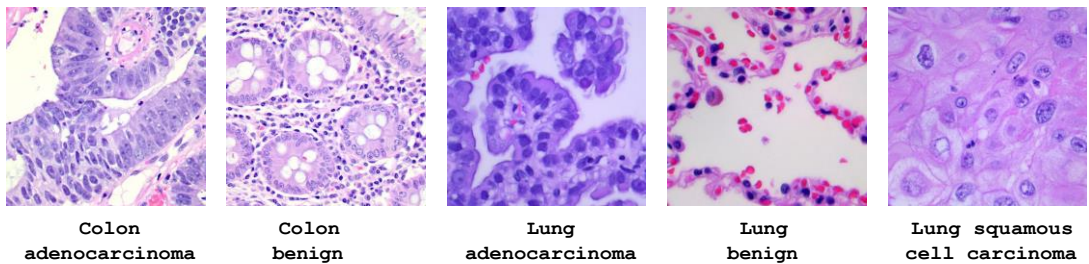

Fig. 1. Lung and Colon Cancer Sample Image Classes

One example of them is the VGG19 model used in [5]. The researchers obtained excellent accuracy but faced problems with computing efficiency, limiting its practical use. R. R. Swain et al. [6] proposed an approach for the classification of colon cancer by using different machine learning algorithms like Random Forests, Decision Tree, Support vector Machine, Naive Bayes, and K-Nearest Neighbour. They have achieved a 95% accuracy by utilizing their algorithms. S. Mangal et al. [1] suggested a convolutional neural network (CNN) for the diagnosis of lung and colon cancer. Their proposed model is a shallow neural network architecture, and by utilizing this network on the LC25000 dataset, they achieved 96% accuracy for colon and 97% accuracy for lung cancer diagnosis. M. Masud et al. [7] developed a framework based on deep learning and image processing techniques. This framework can differentiate five classes of lung and colon cancer from histopathological images with a maximum accuracy of 96.33% on test image data. M. Al-Mamun Provath et al. [8] suggested a new method for maintaining computational efficiency by utilizing the concept of cyclic learning rate. They also used the LC25000 histopathological image-based dataset. The authors employed a CNN model with the proposed methodology, and the suggested model achieves 97% accuracy.

S. Wadekar et al. [9] proposed a modified convolutional neural network by using the pre-trained VGG19 model with tuning procedures. They have also applied the augmentation method to the data and achieved an accuracy of 97.73%. In [10], the authors proposed a deep learning approach that employed a CNN model, where the model comes from the VGG16 model architecture. They applied the Contrast Limited Adaptive Histogram Equalization (CLAHE) technique to enhance the image quality of the dataset. They showed that their proposed model with the CLAHE method can perform with a maximum accuracy of 98.96%. A. Sultana et al. [11] proposed a strategy that contains the study of the comparison of different transfer learning models and a hybrid CNN-SVM model for classifying the lung cancer of three classes. Among all the models, the Inception-V3 model can perform 99.13% accuracy. M. A. Hasan et al. [12] suggest a CNN model based on the power of deep learning methods. Their proposed model architecture is a lightweight model due to its low number of parameters as well as multi-scale mechanism. By using the multiscale concepts, the proposed model can extract local and global features from the images, and they achieved 99.20% accuracy. Study [3] shows the power of residual attention mechanisms in the medical image classification task. This type of model can outperform the state-of-the-art performance in image classification tasks.

Due to the purpose of utilizing the power of the residual attention mechanism-based model, we have applied a modified version of the residual attention model to our utilized dataset that also outperformed compared to some existing.

## III. MATERIALS AND METHODOLOGY

The following section presents the utilized dataset in our research and outlines the complete proposed methodology that we have accomplished.

### A. Dataset Description

For the purpose of this entire study, a dataset of 25,000 high-resolution histopathological images (768 × 768 pixels, JPEG format) across two, three, and five classes for colon cancer, lung cancer, and colon and lung cancer data, respectively. The dataset originated from 1,250 HIPAA-compliant images, comprising 750 lung tissue images (250 each of benign, adenocarcinoma, and squamous cell carcinoma) and 500 colon tissue images (250 each of benign and adenocarcinoma). Using the Augmentor package, the original set was expanded to 25,000 images [13]. Some image samples are shown in Fig. 1. This augmentation process resulted in a balanced distribution across five classes, each containing 5,000 images. The augmentation enhanced dataset robustness while preserving original tissue characteristics, providing a comprehensive foundation for machine learning models in the analysis of histopathological images, particularly for lung and colon cancer detection and classification.

### B. Proposed Methodology

The flowchart of our proposed methodology is illustrated in Fig. 2.

*1) Data Pre-processing:* In this step, we resized the images to 224 × 224 pixels. Subsequently, we employed the min-max normalization technique to scale the values of the pixels of images within a defined range, often between 0 and 1, thereby enhancing the stability and efficiency of the learning process in deep learning models. In histopathological images, min-max normalization reduces the effect of illumination variation and staining differences, making the model focus on the actual tissue structures rather than intensity discrepancies, which can improve model performance [14]. The formula for min-max normalization is as follows:

$$X_{min-max} = \frac{X - X_{\min}}{X_{\max} - X_{\min}} \quad (1)$$

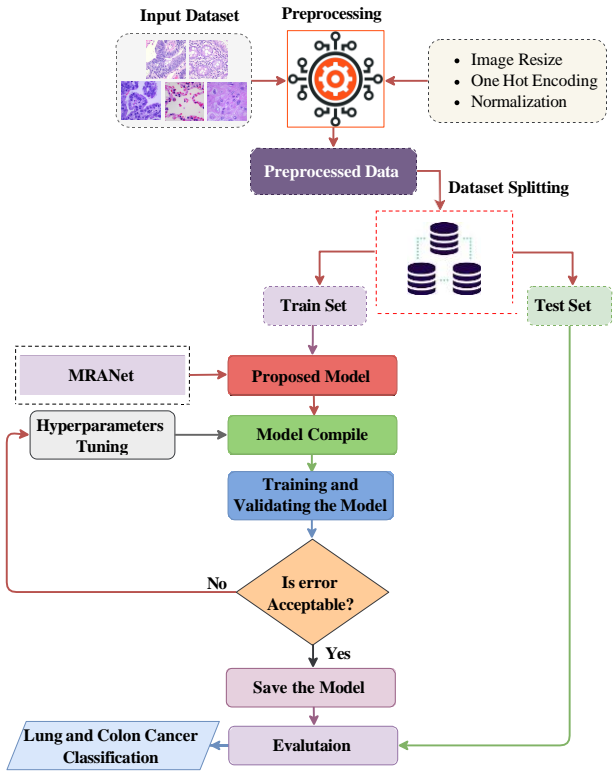

Fig. 2. Schematic Representation of Proposed Methodology

Where, $X$ is the original pixel value, and $X_{\min}$ and $X_{\max}$ are the minimum and maximum pixel values in the image, respectively.

*2) Proposed Network Architecture:* The proposed model is an adaption of the Residual Learning Network and attention mechanism, refined for the classification of lung and colon cancer subtypes. Fig. 3 illustrates the architectural layout of the suggested model.

The proposed design consists of a series of attention modules organized in layers. Each module consists of two separate branches: the mask branch and the trunk branch. The trunk branch does feature selection, while the Mask branch is employed for down-sampling to create a soft mask. Upon concluding the module, the outcomes from both branches are amalgamated and transmitted to the succeeding module. This operation is performed repeatedly. In Fig. 3, $p$ denotes the quantity of pre-processing residual units, $t$ represents the number of residual units in the trunk branch, and $r$ indicates the count of residual units located between the mask branch and the pooling layer.

**Trunk Branch:** The upper branch of the attention module is identified. As mentioned above, its purpose is the extraction of the attention-aware features [3]. For any given input $x$, let its output be denoted by $T(x)$.

**Mask Branch:** The approach employs a bottom-up and top-down structure to learn the mask $M(x)$. The function $M(x)$ serves as control gates analogous to those utilized in LSTM networks or the highway network discussed in the pertinent literature section [3].

The attention module's output can be articulated as follows:

$$H_{i,c}(x) = M_{i,c}(x) * T_{i,c}(x) \quad (2)$$

During backpropagating:

$$\frac{\partial M(x,\theta) T(x,\phi)}{\partial \phi} = M(x,\theta) \frac{\partial T(x,\phi)}{\partial \phi} \quad (3)$$

In this context, $\theta$ represents the parameters of the mask branch, while $\phi$ denotes the parameters of the trunk branch. The attention mask functions as a filter for gradient updates in the backward pass. The design of the mask branch prevents the backpropagation of noisy gradients from incorrect labels, thereby safeguarding the parameter updates of the trunk branch.

**Attention Residual Learning:** The authors state that stacking of Attention modules results in an improvement in performance; however, there is a caveat. Naively stacking layers may be detrimental, as the consecutive dot products between layers with $M(x)$ taking a value between 0 and 1 cause the vanishing of the feature value with increasing depth. To circumvent this problem, the soft mask is modeled as a residual connection [3], described by the following equation:

$$H_{i,c}(x) = (1 + M_{i,c}(x)) * F_{i,c}(x) \quad (4)$$

As a result, the original features $F(x)$ of the trunk branch are preserved, and stacking consecutive modules improves the performance as the feature maps become progressively more refined and clearer with an increase in the number of layers.

**Soft Mask Branch:** A schematic representation of the soft mask branch is shown in Fig. 4, and this branch is inspired by the stacked hourglass network for human pose estimation in that it combines global information obtained from a feed-forward pass step and local pixel-level information from the top-down feedback steps. Max pooling is applied repeatedly to the input image to downsample it until the lowest resolution is reached. Following this, linear interpolation layers upsample the output of the max pooling layers. Skip connections between these parts of the network allow both top-level (global) and bottom-level (local) data to influence the training. Finally, a sigmoid layer is applied to normalize the outputs to lie in between 0 and 1 after two 1x1 convolutions [3].

*C. Environment Setup and Training Procedure*

This section outlines the environment setup and training procedure. The experiments were conducted in a high-performance computing environment using the Google Cloud environment with one GPU (NVIDIA Tesla P100) and the TensorFlow (2.0) and TensorBoard (2.0.2) libraries. In our study, we employed the utilized dataset split of 60% for training, 20% for validation, and 20% for testing. This distribution ensured a robust training process while reserving sufficient data for model validation and final performance evaluation. For the purpose of training, we use the Adam optimizer [15] with a mini-batch size of 32. Additionally, we also applied the Adam optimizer learning rate to 0.0001 and used the categorical cross-entropy loss function with the epoch number of 50. To improve model performance, two different regularization methods are applied, such as batch normalization and dropout [16], to avoid overfitting and improve the speed, performance, and stability of our models.

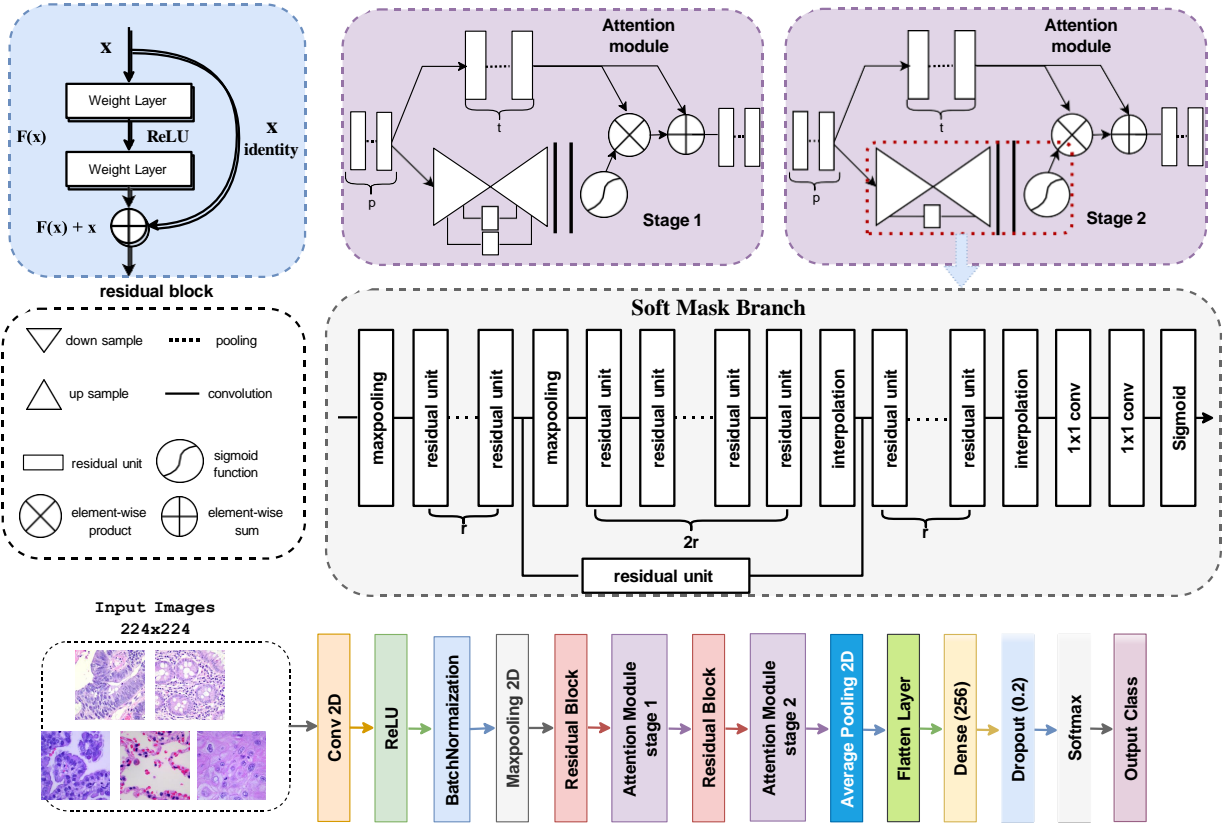

Fig. 3. Schematic Representation of Proposed Model Architecture

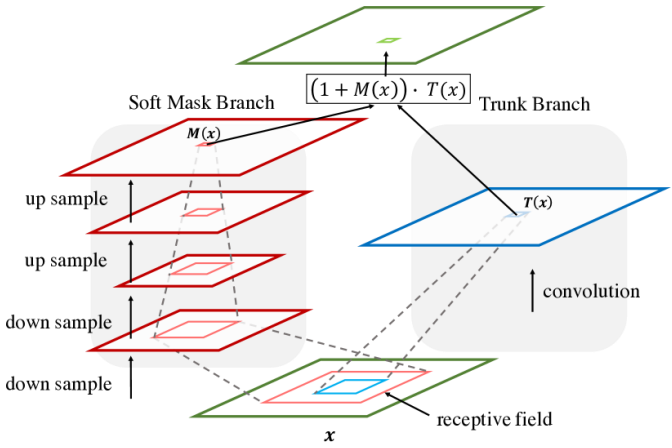

Fig. 4. Attention module with trunk branch and mask branch

## IV. RESULTS AND DISCUSSION

After successfully training our model with the dataset, we have evaluated our model on test data that the model has not seen before through several evaluation metrics such as precision, sensitivity, specificity, accuracy, Matthews correlation coefficient (MCC), f1-score, Kappa-score, and AUC-ROC. We have performed our experiments on two, three, and five-class datasets to check our model's efficiency. The obtained results through our experiments are indexed in Table I.

Table I depicts the performance analysis of the proposed model for different classes. Our proposed MRANet model demonstrated superior performance compared to other state-of-the-art architectures in this classification task. The table provides a performance analysis of a proposed model across varying numbers of classes (2, 3, and 5). The model achieves exceptional performance for 2 classes, with a precision of 99.90%, sensitivity of 98.72%, specificity of 99.90%, accu-racy of 99.30%, MCC (Matthews Correlation Coefficient) of 98.61%, F1-score of 99.30%, kappa-score of 98.60%, and AUC-ROC of 0.993. For 3 classes, the metrics show a slight decline but remain robust, with a precision of 94.40%, sensitivity of 95.45%, specificity of 97.22%, accuracy of 96.63%, MCC of 92.41%, F1-score of 94.92%, kappa-score of 94.85%, and AUC-ROC of 0.974. For 5 classes, the model demonstrates consistent performance, with a precision of 94.60%, sensitivity of 93.29%, specificity of 98.65%, accuracy of 97.56%, MCC of 92.42%, F1-score of 93.94%, kappa-score of 96.47%, and AUC-ROC of 0.984. Overall, the model exhibits high reliability, maintaining strong classification performance across all metrics and class distributions.

TABLE I
PROPOSED MODEL EVALUATION METRICS RESULT FOR DIFFERENT CLASSES

| Number of Class | Precision (%) | Sensitivity (%) | Specificity (%) | Accuracy (%) | MCC (%) | F1-Score (%) | Kappa-Score (%) | AUC-ROC |
| --- | --- | --- | --- | --- | --- | --- | --- | --- |
| 2 Classes | 99.90 | 98.72 | 99.90 | 99.30 | 98.61 | 99.30 | 98.60 | 0.993 |
| 3 Classes | 94.40 | 95.45 | 97.22 | 96.63 | 92.41 | 94.92 | 94.85 | 0.974 |
| 5 Classes | 94.60 | 93.29 | 98.65 | 97.56 | 92.42 | 93.94 | 96.47 | 0.984 |

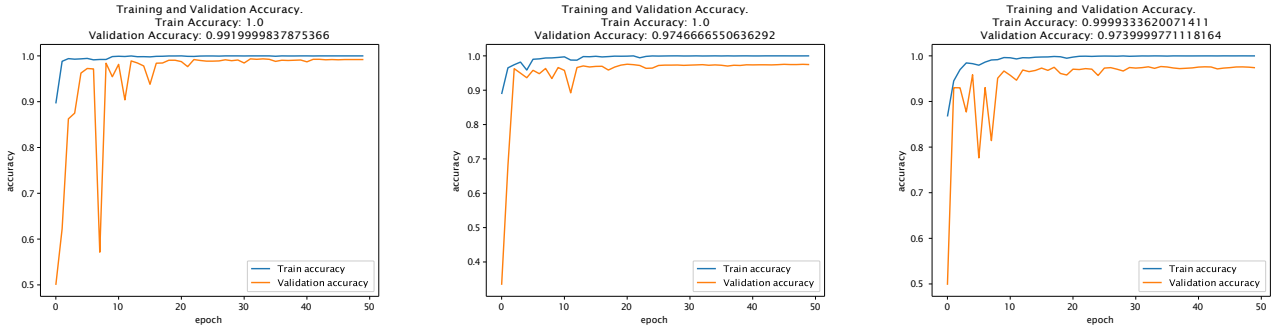

Fig. 5. Proposed model training and validation accuracy curve for different classes

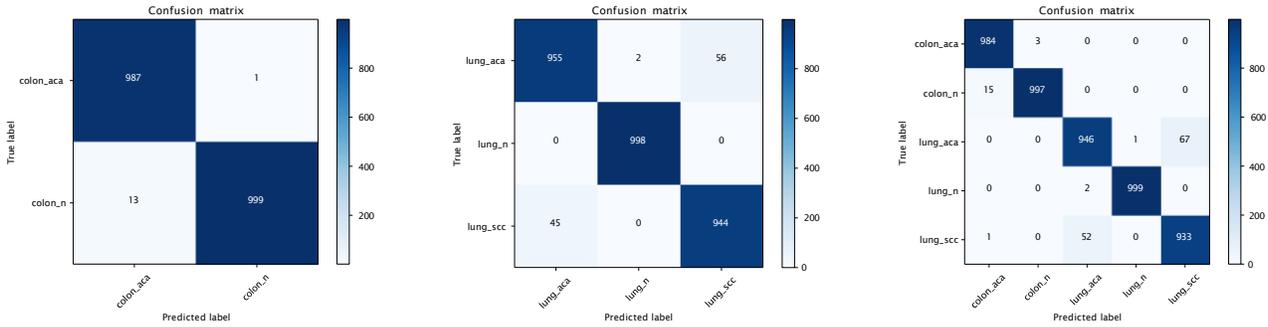

Fig. 6. Proposed model confusion matrix for different classes

The provided Fig. 5 depicts the training and validation accuracy trends for the proposed model across 2, 3, and 5 classes. The model achieves perfect training accuracy of 100% for classes 2 and 3 and near-perfect validation accuracy of 99.19%, and 97.47% respectively, indicating exceptional learning and generalization capabilities with minimal overfitting. For 5 classes, the training accuracy is 99.93%, while the validation accuracy is 97.39%, reflecting excellent learning and generalization with slightly higher complexity. Across all cases, the training and validation accuracy curves converge smoothly, highlighting the model's stability and efficient optimization. The consistent performance observed across various class configurations highlights the model's adaptability and effectiveness.

Fig. 6 illustrates the confusion matrices of the proposed model. The model demonstrates excellent performance with only minor misclassifications for binary classification. For 3 classes, the confusion matrix shows high precision and recall, with most predictions correct, except for 56 instances of "lung_aca" misclassified as "lung_scc", and 45 "lung_scc" misclassified as lung_aca. The performance remains robust for 5 classes, with a slight increase in misclassifications due to the higher complexity. Here, 67 "lung_aca" instances are misclassified as "lung_scc" and 52 "lung_scc" instances are incorrectly predicted as "lung_n". Overall, the confusion matrices confirm the model's strong classification ability across all scenarios, with high accuracy and minimal misclassifica-tions despite increasing class complexity.

TABLE II
COMPARISON WITH EXISTING LITERATURE

| Authors | Methods | Accuracy(%) |
|---|---|---|
| R. R. Swain et al. [6] | ML algorithms | 95% |
| S. Mangal et al. [1] | CNN | 96% |
| M. Masud et al. [7] | Deep Learning Model | 96.33% |
| M. Al-Mamun Provath et al. [8] | CNN | 97% |
| S. Wadekar et al. [9] | VGG16 | 97.73% |
| S. Hadiyoso et al. [10] | VGG16 with CLAHE | 98.96% |
| A. Sultana et al. [11] | Hybrid CNN-SVM | 99.13% |
| M. A. Hasan et al. [12] | Lightweight CNN | 99.20% |
| Proposed | MRANet | 99.30% 96.63%, 97.56% |

Overall, the proposed model showcases strong adaptability, generalization, and reliability across varying class distributions. The improved performance of the proposed model is attributed to the modified residual attention network, which combines attention mechanisms and residual connections. The attention modules enhance feature extraction by focusing on relevant data, while residual connections improve gradient flow and enable deeper network training. This architecture captures complex patterns effectively, ensuring robust performance across varying class complexities. Additionally, the design ensures training stability and efficient optimization, leading to superior accuracy, sensitivity, and generalization.

## V. CONCLUSION

In this research, we introduced an innovative method for classifying lung and colon cancers through histopathological images, attaining a cutting-edge accuracy of 99.30%, 96.63%, and 97.56% for two, three, and five classes. Our proposed model, based on a modified residual attention network architecture, outperformed existing methods in terms of several metrics. The novelty of our research lies in how to improve the classification accuracy in different numbers of classes. Our model's ability to accurately distinguish between different cancer subtypes and benign tissues represents a significant advancement in automated lung and colon cancer diagnosis.

Future efforts should aim to enhance the model's capabilities to encompass a wider variety of cancer types and stages. Our proposed model can be integrated into a real-time lung and colon cancer detection system for the purpose of using clinicians, and we also developed a mobile-based application for developing countries to assist in diagnosing and clinical practice.

## ACKNOWLEDGMENT

The authors thankfully acknowledge the organization and authors for developing the datasets for early diagnosis of lung and colon cancer classification.## REFERENCES

[1] S. Mangal, A. Chaurasia, and A. Khajanchi, "Convolution neural networks for diagnosing colon and lung cancer histopathological images," *arXiv preprint arXiv:2009.03878*, 2020.

[2] N. Shahadat, R. Lama, and A. Nguyen, "Lung and colon cancer detection using a deep ai model," *Cancers*, vol. 16, no. 22, p. 3879, 2024.

[3] F. Wang, M. Jiang, C. Qian, S. Yang, C. Li, H. Zhang, X. Wang, and X. Tang, "Residual attention network for image classification," in *Proceedings of the IEEE conference on computer vision and pattern recognition*, 2017, pp. 3156–3164.

[4] X. Li, L. Zhang, J. Yang, and F. Teng, "Role of artificial intelligence in medical image analysis: A review of current trends and future directions," *Journal of Medical and Biological Engineering*, pp. 1–13, 2024.

[5] B. L. Qasthari, E. Susanti, and M. Sholeh, "Classification of lung and colon cancer histopathological images using convolutional neural network (cnn) method on a pre-trained models," *International Journal of Applied Sciences and Smart Technologies*, vol. 5, no. 1, pp. 133–142, 2023.

[6] R. R. Swain, D. S. K. Nayak, and T. Swarnkar, "A comparative analysis of machine learning models for colon cancer classification," in *2023 International Conference in Advances in Power, Signal, and Information Technology (APSIT)*. IEEE, 2023, pp. 1–5.

[7] M. Masud, N. Sikder, A.-A. Nahid, A. K. Bairagi, and M. A. AlZain, "A machine learning approach to diagnosing lung and colon cancer using a deep learning-based classification framework," *Sensors*, vol. 21, no. 3, p. 748, 2021.

[8] M. Al-Mamun Provath, K. Deb, and K.-H. Jo, "Classification of lung and colon cancer using deep learning method," in *International Workshop on Frontiers of Computer Vision*. Springer, 2023, pp. 56–70.

[9] S. Wadekar and D. K. Singh, "A modified convolutional neural network framework for categorizing lung cell histopathological image based on residual network," *Healthcare Analytics*, vol. 4, p. 100224, 2023.

[10] S. Hadiyoso, S. Aulia, I. D. Irawati et al., "Diagnosis of lung and colon cancer based on clinical pathology images using convolutional neural network and clahe framework," *International Journal of Applied Science and Engineering*, vol. 20, no. 1, pp. 1–7, 2023.

[11] A. Sultana, T. T. Khan, and T. Hossain, "Comparison of four transfer learning and hybrid cnn models on three types of lung cancer," in *2021 5th International Conference on Electrical Information and Communication Technology (EICT)*. IEEE, 2021, pp. 1–6.

[12] M. A. Hasan, F. Haque, S. R. Sabuj, H. Sarker, M. O. F. Goni, F. Rahman, and M. M. Rashid, "An end-to-end lightweight multi-scale cnn for the classification of lung and colon cancer with xai integration," *Technologies*, vol. 12, no. 4, p. 56, 2024.

[13] A. A. Borkowski, M. M. Bui, L. B. Thomas, C. P. Wilson, L. A. DeLand, and S. M. Mastorides, "Lung and colon cancer histopathological image dataset (lc25000)," *arXiv preprint arXiv:1912.12142*, 2019.

[14] S. Patro, "Normalization: A preprocessing stage," *arXiv preprint arXiv:1503.06462*, 2015.

[15] D. P. Kingma, "Adam: A method for stochastic optimization," *arXiv preprint arXiv:1412.6980*, 2014.

[16] N. Srivastava, G. Hinton, A. Krizhevsky, I. Sutskever, and R. Salakhutdinov, "Dropout: a simple way to prevent neural networks from overfitting," *The journal of machine learning research*, vol. 15, no. 1, pp. 1929–1958, 2014.